\documentclass[aps,pre,twocolumn,superscriptaddress,floatfix,amsfonts,citeautoscript]{revtex4-1}
\usepackage{times}
\usepackage{graphicx}
\usepackage{graphics}
\usepackage{dcolumn}
\usepackage{bm}
\usepackage{color}
\usepackage{float}
\usepackage{amssymb} 
\usepackage{amsmath}
\usepackage{multirow}
\begin{document}

\title{\textbf{Analysis of vibrational normal modes for Coulomb clusters}}

\author{Biswarup Ash}
\thanks{Present address: Department of Condensed Matter Physics, Weizmann Institute of Science, Rehovot 76100, Israel}
\affiliation{Indian Institute of Science Education and Research-Kolkata, Mohanpur, India-741246}

\author{Chandan Dasgupta}
\affiliation{Department of Physics, Indian Institute of Science, Bangalore 560012, India}
\affiliation{International Centre for Theoretical Sciences, TIFR, Bangalore 560012, India}

\author{Amit Ghosal}
\affiliation{Indian Institute of Science Education and Research-Kolkata, Mohanpur, India-741246}

\begin{abstract}
We study various properties of the vibrational normal modes for Coulomb-interacting particles in two-dimensional irregular confinement using numerical simulations. By analyzing the participation ratio and spectral statistics, we characterize the vibrational modes for Coulomb clusters as localized, quasi-localized and delocalized. We also study a novel correlation function to understand the spatial structure of these different kinds of modes and subsequently extract the associated characteristic length scales. We further demonstrate that, at any given temperature, particles exhibiting larger displacement over a time interval comparable to the structural relaxation time, are strongly correlated with the low-frequency quasi-localized modes of the inherent structure corresponding to the initial configuration. Establishing this correlation for Coulomb clusters paves the path to identify the particular feature of the initial configuration that determines the previously observed heterogeneous dynamics of the particles at low temperatures in these systems. 
\end{abstract}

\maketitle

\section{INTRODUCTION} 

Finite systems with interacting particles are of fundamental interest as they bridge the gap between the intriguing phenomenology of a single particle in confinements and complex many particle effects in bulk (extended) systems. A systematic study of the properties of finite systems can help in understanding the intricacies of inter-particle interactions and boundary effects. Theoretical and experimental studies have shown that depending on the geometry of the confinement and the nature of the inter-particle interactions, particles in finite systems exhibit different structural and motional signatures~\cite{Exp_melzer, PhysRevE.91.012305, Ptr_anistrpy_log, Yurtsever,PRE17}. Disorder, which is intrinsic to all real materials, can also be introduced in the nanoclusters, primarily, through the irregularities in the geometry of the confinements. Additionally, greater experimental tunability for finite systems compared to its bulk counterpart makes these systems an ideal playground for exploring the complex interplay of disorder and interaction~\cite{Chui,Marcus_QD,VD2,DA13,DA16}. 

Among various finite systems, Coulomb interacting particles in traps, the finite size analogs of Wigner crystals~\cite{Wigner1934}, have drawn considerable theoretical~\cite{BP94,Peeters_NJP,Bonitz_2008,DA13,DA16,EPL2016,PRE17} and experimental~\cite{He_wigner, QD_Ashoori, QD,RF_trap, PhysRevE.67.016411, Melzer_INM} attention in last few decades. Extensive studies of the static and dynamic properties of Coulomb interacting particles in parabolic confinements across a wide range of temperatures $(T)$ have established a thermal crossover from solid to liquid like phases. In these systems, when the number of particles, $N$, is small, it has been found that particles arrange themselves in concentric circular rings (shell structure) in the ground state $(T=0)$. A Mendeleev-type periodic table is also made for the arrangement of the particles in parabolic trap~\cite{BP94}. On the other hand, for large systems $(N > 200)$, a triangular lattice structure appears in the central region of the parabolic confinement and circular ring structures are found for the particles near the boundary. Further it has also been established that such shell-structured systems melt typically in two-steps~\cite{BP94}: The first transition (at a lower $T$) corresponds to inter-shell rotations, and the second one, occurring at a higher $T$, corresponds to inter-shell diffusion where particles wander freely within the shells and can also hop from one shell to the other. Along with the melting, properties of the vibrational spectra for Coulomb interacting particles in parabolic confinement were also looked into in great detail~\cite{Peeters_NM}.

Melting in finite systems are mostly studied in confinements having circular symmetry where one can exploit this symmetry to identify the signatures of melting. So, it is natural to ask how does the melting scenario get altered in the absence of circular symmetry? Recently this question has been addressed by studying the static~\cite{DA13} and dynamic~\cite{EPL2016,PRE17} properties of Coulomb interacting particles in an irregular confinement which breaks all spatial symmetries. It is shown that while the positional order is highly depleted even at the lowest temperature, a solid like phase can still be identified from the presence of strong bond-orientational order~\cite{DA13,PRE17}. With increasing temperature, such an orientationally ordered `solid' crosses over to a disordered liquid like phase. From the temperature dependence of several observables such as Lindemann ratio, specific heat~\cite{DA13}, and generalized susceptibilities~\cite{PRE17}, a more-or-less unique crossover temperature $(T_X)$ is also identified. 

Study of the dynamic properties of Coulomb interacting particles in irregular confinement revealed the possibility of observing the key signatures of glassy dynamics in the context of finite systems with long-range interacting particles~\cite{EPL2016,PRE17}. Extensive molecular dynamics simulations showed that the particles exhibit spatially correlated motion at low temperature $(T<T_X)$ and consequently, the distribution of the displacement of particles becomes non-Gaussian~\cite{EPL2016}. Particularly, the tail of the distribution of the displacement of particles turned out to be exponential for small $T (<T_X)$ and stretched exponential (decay slower than exponential) for $T\sim T_X$ at intermediate and long times (compared to the structural relaxation time)~\cite{EPL2016,PRE17}. Analysis of the trajectory of individual particles at low $T$ revealed that dynamics of the particles is strongly heterogeneous: while many particles execute small amplitude vibrational motion around their equilibrium positions for long times, other particles become highly mobile, carrying out longer displacements (compared to the average inter-particle distance) in the same time scale. Though it is found that these mobile particles form tortuous string-like paths, their locations appear to have no preference for the bulk or the boundary~\cite{EPL2016}. 

The consequences of the heterogeneous dynamics in irregular Coulomb clusters were recently addressed~\cite{EPL2016,PRE17}, but there are still open questions: Why do some particles become highly mobile compared to others? Is the appearance of mobile particles in certain spatial locations completely random? Is there any characteristics of the system that can help to predict which particle would exhibit larger displacement at long time? The main objective of this paper is to address these questions by investigating possible connections between the properties of low-lying local minima in the energy landscape (called `inherent structures', as we will see) and the dynamic heterogeneity developed in the system after a long time. Establishing any such correlation would certainly help in understanding the observed heterogeneous dynamics of the particles in Coulomb clusters.  

Heterogeneous dynamics at low temperatures is quite ubiquitous in disordered systems such as supercooled liquids~\cite{Ediger_2000}. For glassy dynamics, one of the central issues is to understand the structural origin of the slow heterogeneous dynamics. It has been found that analysis of the vibrational normal modes, which encode information on how each particle proposes to move when all of them undergo collective motion, can be a fruitful approach in identifying the structural origin of dynamical heterogeneity~\cite{Widmer_Natphys,Widmer_JCP,Manning,Brito07, PhysRevLett.105.025501, PhysRevLett.104.248305,PhysRevLett.107.188303}.

Recent computer simulation studies of supercooled liquids suggest that the spatial regions where particles are more susceptible to experience longer displacement result from low-frequency quasi-localized modes, known as `soft glassy modes', of the system~\cite{Widmer_Natphys,Widmer_JCP,Brito07,Manning}. The existence, as well as the nature of such soft modes, is a subject of much current interest as they are believed to be  intimately related to the anomalous low-temperature properties of amorphous systems~\cite{Franz24112015,MizunoE9767,Mizuno11949,Ikeda2,Barrat2,Manning}. Using normal mode analysis, their existence has been reported recently in experiments on colloidal glasses~\cite{PhysRevLett.104.248305,PhysRevLett.104.248305, PhysRevLett.107.188303,C0SM00265H}. Since, the dynamical features of Coulomb interacting particles in irregular confinement have resemblance with those of glassy dynamics, we ask: Is there any connection between the low frequency normal modes and the observed heterogeneous dynamics in Coulomb clusters? With this aim, in this work, we study the vibrational (or quenched) normal modes, as well as instantaneous normal modes, of Coulomb interacting particles in irregular confinement, using computer simulation. Since long-wavelength phonon-like modes are not present in small systems with irregular boundary, complications arising from the presence of both phonon-like modes and soft glassy modes in the same frequency range~\cite{lerner_prl,lerner-bouchbinder,MizunoE9767} are not present in the system studied here.

The main outcomes of our analysis can be summarized as follows: Analysis of the vibrational modes in terms of participation ratio and concepts of random matrix theory helps to characterize the normal modes in broadly three different classes: localized mode, quasi-localized mode and delocalized modes. The existence of these quasi-localized modes in disordered solids has been proposed in the literature~\cite{Schober1,Schober2,Widmer_Natphys,Manning,Ikeda2}, but we identify them in this present study for the first time (to the best of our knowledge) in a comprehensive manner in a confined system with irregular boundary. Further, an introduction of a spatial correlation function allow us to identify the typical length scale associated with the quasi-localized modes. We find that the particles with large magnitude of polarization vectors in the low frequency quasi-localized modes subsequently exhibit larger mobility over a long time. We also analyze the density of states and participation ratio for instantaneous normal modes which contain both stable and unstable modes. From the temperature dependence of the fraction of unstable modes, we estimate the crossover temperature for the system and find a good agreement with the previously reported value.          

The rest of the paper is organized as follows: In Sec.~\ref{Sec:Model}, we discuss the details of the models and methods used in our study. In sec.~\ref{sec:VDOS}, we analyze the vibrational modes for our model system in terms of density of states, participation ratio and tools of random matrix theory. On the basis of these quantities, we classify the normal modes in localized, quasi-localized and delocalized modes. We also introduce a spatial correlation function that helps to extract the characteristic length scales associated with different kind of modes. In sec.~\ref{Sec:dispcorln}, we show that a small subset of the low-frequency quasi-localized modes associated with a given configuration can give a good description of the particles that exhibits large displacements at longer times. In sec.~\ref{INM}, we discuss the instantaneous normal modes for the Coulomb clusters. Finally, we conclude in sec.~\ref{sec:conclude}.

\section{Model and Methods}\label{Sec:Model}

We consider $N$ classical particles with Coulomb interaction, each having charge $q$ and trapped by an irregular confinement $V_{\rm conf}^{\rm Ir}$. Particles are restricted to move in two spatial dimension (say, $x$-$y$ plane). The potential energy part of the Hamiltonian for such a system, in dimensionless form, reads as follows:
\begin{equation}
\mathcal{H}= \sum_{i<j=1}^{N} \frac{1}{|\vec{r}_i - \vec{r}_j |} + \sum_{i=1}^{N} V_{\rm conf}^{\rm Ir}(x_i,y_i)
\label{Eq:Hamiltonian}
\end{equation}
where, $|\vec{r}_i|=\sqrt{x_i^2 + y_i^2}$ is the distance of the $i$-th particle at the location $(x_i,y_i)$ from the centre of the confinement. In writing Eq.~\ref{Eq:Hamiltonian}, we set the unit of energy as $E_0=q^2(4\pi\epsilon \xi)^{-1}=1$ where we introduce the unit of length as $\xi=q^2(4\pi\epsilon)^{-1}$ and $\epsilon$ represents the dielectric constant of the medium. The first term in the Hamiltonian stands for the potential energy due to the Coulomb repulsion between the particles. The second term in the Hamiltonian describes the energy due to the two dimensional irregular confinement potential which has the form
\begin{equation}
V_{\rm conf}^{\rm Ir}(x,y)=a\left[ x^4/b+by^4-2\lambda {x^2}{y^2} +\gamma (x-y)xyr\right]. 
\label{Eq:IWM}
\end{equation}
$V_{\rm conf}^{\rm Ir}$ is defined through four parameters; $a,b, \lambda$, and $\gamma$. The overall multiplicative factor $a$, which makes the confinement deep or shallow, controls the average particle density. The parameter $b=\pi/4$ breaks the $x$-$y$ symmetry, $\lambda$  controls the chaotic nature of the dynamics of a particle in this potential and $\gamma$ breaks the reflection symmetry~\cite{Bohigas93,Hong03}. By tuning $\lambda$ from zero to unity, one can generate periodic to chaotic motion for a single particle in the trap~\cite{Bohigas93}. In our study, chaotic dynamics along with the broken spatial symmetries are considered as the footprints of disorder. We consider $\lambda \in [0.565,0.635]$ and $\gamma \in [0.10,0.20]$~\cite{Hong03}. The values of the two parameters $\lambda$ and $\gamma$ are chosen in the above mentioned range to generate self-similar copies of motional signatures in the system. This allow us to collect statistics on the quantities of our interest over those `realizations of disorder', each identified by a specific combination $(\lambda, \gamma)$, for the purpose of ``disorder averaging"~\cite{Ullmo03}. Note that the parameter $a$ is expressed as $a\rightarrow q^2(4\pi\epsilon)^{-1} \xi^{-5} a$ to ensure that all the parameters, $\{a, b, \lambda,\gamma\}$ become dimensionless. 

Note that, we are interested in understanding the generic features of disordered systems. Thus, it is important to take special care in choosing the appropriate confining potential, so that we can capture universal behavior due to disorder. A large number of existing literatures show that our choice of $V_{\rm conf}^{\rm Ir}$, given in Eq.~\ref{Eq:IWM}, actually replicates the universal behavior of generic disordered systems in the chosen range of parameter~\cite{Hong03, Ghosal_PRB05, Ullmo03}. 

To generate the equilibrium configurations for Coulomb interacting particles in irregular confinement at different temperatures $(T)$, we have carried out molecular dynamics (MD) simulation~\cite{FrenkelBook}. To achieve a desired $T$, we have used velocity-rescaling method~\cite{FrenkelBook} during the equilibration. After equilibration, we have implemented conventional velocity-Verlet algorithm~\cite{FrenkelBook} (without velocity-rescaling) to integrate the equations of motion. In our rescaled unit, $t=1$ represents the timescale at which the crossover from ballistic to diffusive behavior takes place in particle dynamics, on an average. Thus, a particle senses the presence of others beyond unit time. We have performed MD runs up to $2\times 10^6$ steps with a time step size of $dt = 0.005$, which yields a total time, $ t \sim 100 \tau_{\alpha}$, where $\tau_{\alpha}$ is the structural relaxation time, at the highest temperature. In this work, we consider $N=150$ Coulomb interacting particles in irregular confinement, same as in Ref.~\cite{EPL2016}. While we present below all our results for systems with $N=150$ particles, we verified that our key conclusions survive also in larger systems with $N=500$ particles. This is demonstrated in Appendix B.

For a given configuration, the normal modes are obtained by diagonalizing the matrix of the second derivatives of the potential energy (or the Hessian matrix) with respect to the coordinates of the particles in that configuration~\cite{PhysRevLett.74.936}. The eigenvalues $(\lambda)$ of the Hessian matrix are related to the frequency $(\omega)$ of the normal modes by the relation $\omega=\sqrt{\lambda}$  and the corresponding eigenvectors characterize the normal modes of the configuration under consideration. If the configuration is a representative of an equilibrium state of the system, then the obtained normal modes are called the instantaneous normal modes (INMs). On the other hand, if the configuration is an inherent structure (IS), an energy minimized configuration corresponding to an equilibrium configuration, then the normal modes are called the vibrational or quenched normal modes (QNMs). To obtain the QNMs, we first quench each equilibrium configuration, obtained from MD simulations, to its corresponding IS, using the conjugate gradient method~\cite{Num_Recipe}. Since, the obtained normal modes correspond to an IS, the eigenvalues of the Hessian matrix are all positive.   

\section{Analysis of Vibrational (quenched) modes}\label{sec:VDOS}

At any given temperature, following the procedure described in the last section, we compute the quenched normal modes for an ensemble of ISs. From this collection of modes, we evaluate the the vibrational density of states of the system which represents the probability density for the quenched normal modes of a given frequency.

\begin{figure}
\includegraphics[width=8.9cm,keepaspectratio]{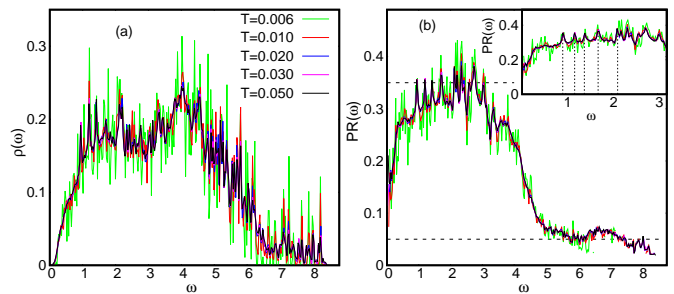}
\caption{ (Color online)(a) Density of states, $\rho(\omega)$, of the quenched normal modes as a function of frequency $\omega$, for $N=150$ particles in irregular confinement for different temperatures. (b) The average participation ratio PR$(\omega)$ as a function of $\omega$ for the spectra shown in (a). Results are obtained by collecting statistics over $6$ independent realizations, each identified by specific combination of irregularity parameters ${\lambda,\gamma}$, as discussed in text. The upper horizontal dotted lines demarcates the boundary between delocalized and quasi-localized modes, while the lower one defines the boundary between quasi-localized and localized modes. Inset of panel (b) shows PR$(\omega)$ for $\omega<3.2$ to emphasize that there are certain modes (indicated by vertical dashed lines) at which PR$(\omega)$ shows peaks that are robust to temperature. 
}
\label{fig:IWM_QNM_DOS_PR}
\end{figure}

\subsection{Density of states and participation ratio}
In Fig.~\ref{fig:IWM_QNM_DOS_PR}(a), we show the temperature dependence of the normalized density of states (DOS), $\rho(\omega)$, where 
\begin{equation}
\rho(\omega) =\left\langle \frac{1}{2N} \sum_{l=1}^{2N} \delta (\omega - \omega_l)\right\rangle,
\end{equation}
by constructing the histogram of the frequencies $\omega$ for the quenched normal modes for $N=150$ particles in irregular confinement. The normalization for DOS is 
\begin{equation}
\int \rho(\omega) \mathrm{d}\omega = 1.
\end{equation}
To improve the statistics, we collect all the QNMs obtained from the available ISs for a given $T$. At low $T$, $\rho(\omega)$ shows a very peaked structure depicting the fact that the system explores only a few distinct inherent structures and thus only certain modes at particular frequencies can occur in this state. For higher temperatures, a more continuous mode spectrum $\rho(\omega)$ is observed.

To characterize the nature of different modes we compute the participation ratio, PR$(\omega)$, for each mode. The participation ratio PR$(\omega)$ quantifies the localization properties of a normal mode by measuring what fraction of particles contributes significantly to a given mode. It is defined as
\begin{equation}
\text{PR}(\omega_l) = \left[ N \sum_{i=1}^{N} \left( \vec{e}_i^{~l} \cdot  \vec{e}_i^{~l} \right)^2 \right]^{-1}
\end{equation}
where $\vec{e}_i^{~l}$ is the contribution of the particle $i$ to the normalized eigenvector $\vec{e}^{~l}$. For a perfectly delocalized mode where all particles contribute equally to the eigenvector, PR is one. Similarly, for an ideal localized mode where only one particle contributes to the eigenvector, PR is $1/N$. Thus, for delocalized modes, PR is of order unity while for localized (or quasi-localized) modes, it will scale inversely with the system size, vanishing in the bulk limit.

Fig.~\ref{fig:IWM_QNM_DOS_PR}(b) shows average participation ratio, where averaging is done by considering all the modes in each histogram bin, as a function of mode frequency, $\omega$, for various spectra shown in Fig.~\ref{fig:IWM_QNM_DOS_PR}(a). We find that  PR$(\omega)$ is smaller for very low and high frequency modes and relatively higher for intermediate frequencies. Thus, very low and high-frequency modes are more localized compared to those at the intermediate frequencies. In Fig.~\ref{fig:IWM_QNM_DOS_PR}(b), we find that there are few robust low-lying modes for which PR$(\omega)$ shows peaks which persist at all temperatures. These are the same modes for which we find peaks robust to $T$ in $\rho(\omega)$ as well (Fig.~\ref{fig:IWM_QNM_DOS_PR}(a)). Note that the maximum value of the average participation ratio $(\sim 0.43)$ is far below unity. Thus, the value of the participation ratio alone is not sufficient to classify the modes at a given frequency as localized or deloaclized.  

Although the participation ratio is relatively higher for modes with intermediate frequencies compared to that for low and high-frequency modes, in order to characterize them as localized, delocalized or even quasi-localized modes, we, next, use the concepts of random matrix theory (RMT), namely the statistics of spacings between the successive eigenvalues of the Hessian matrix~\cite{Mehta_RMT, GUHR1998189}. The elements of the Hessian matrix can be considered as random variables, as these depend on the random positions of the particles. Thus, the Hessian matrix can be treated as a random matrix ensemble~\cite{MEZARD,Clapa} and we, next, use the tools developed in the context of RMT to make clear distinction between the delocalized and localized modes in our system.
\begin{figure}[t!]
\hspace{-0.9cm}\includegraphics[width=7.85cm,keepaspectratio]{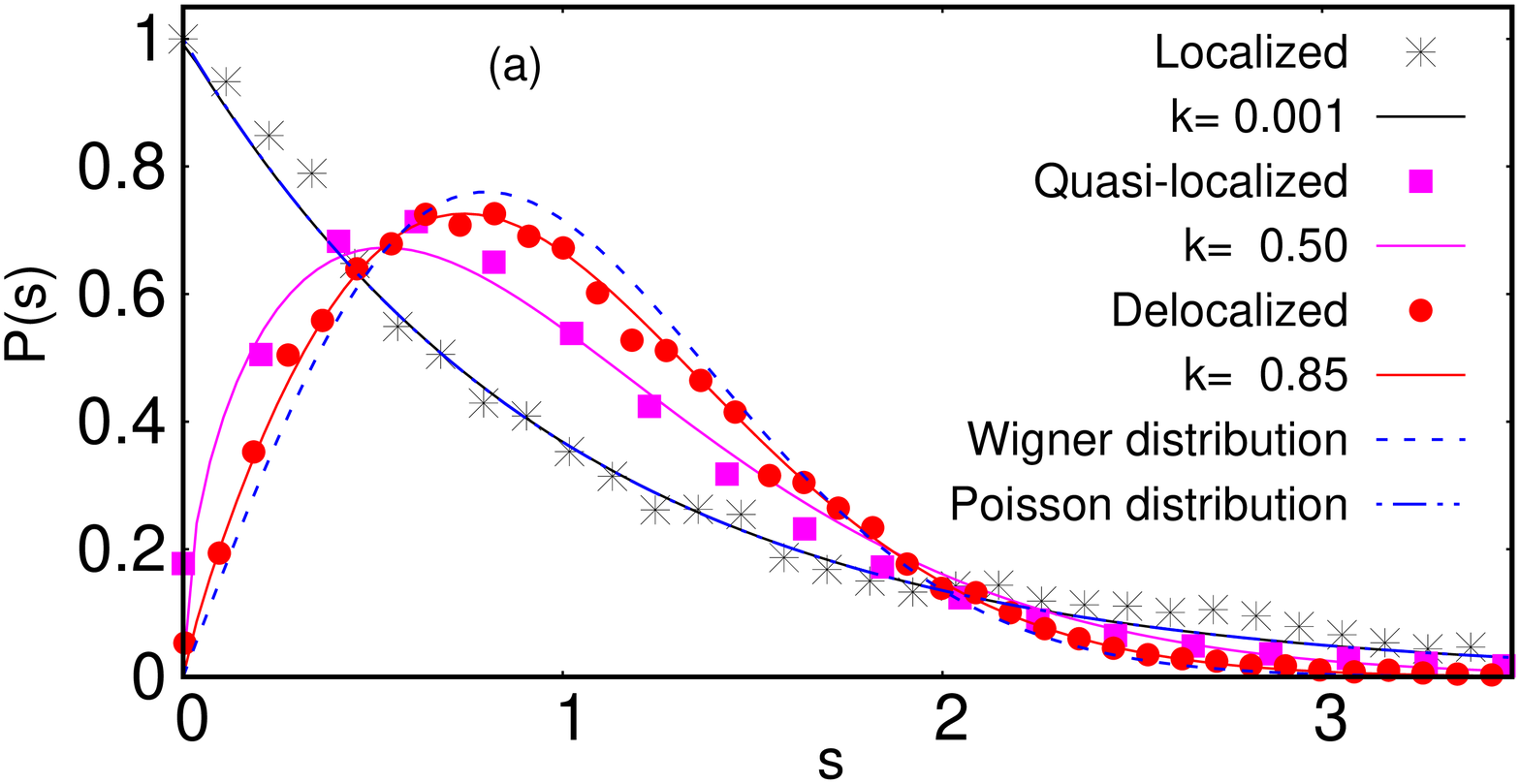}\\
\includegraphics[width=8.0cm,keepaspectratio]{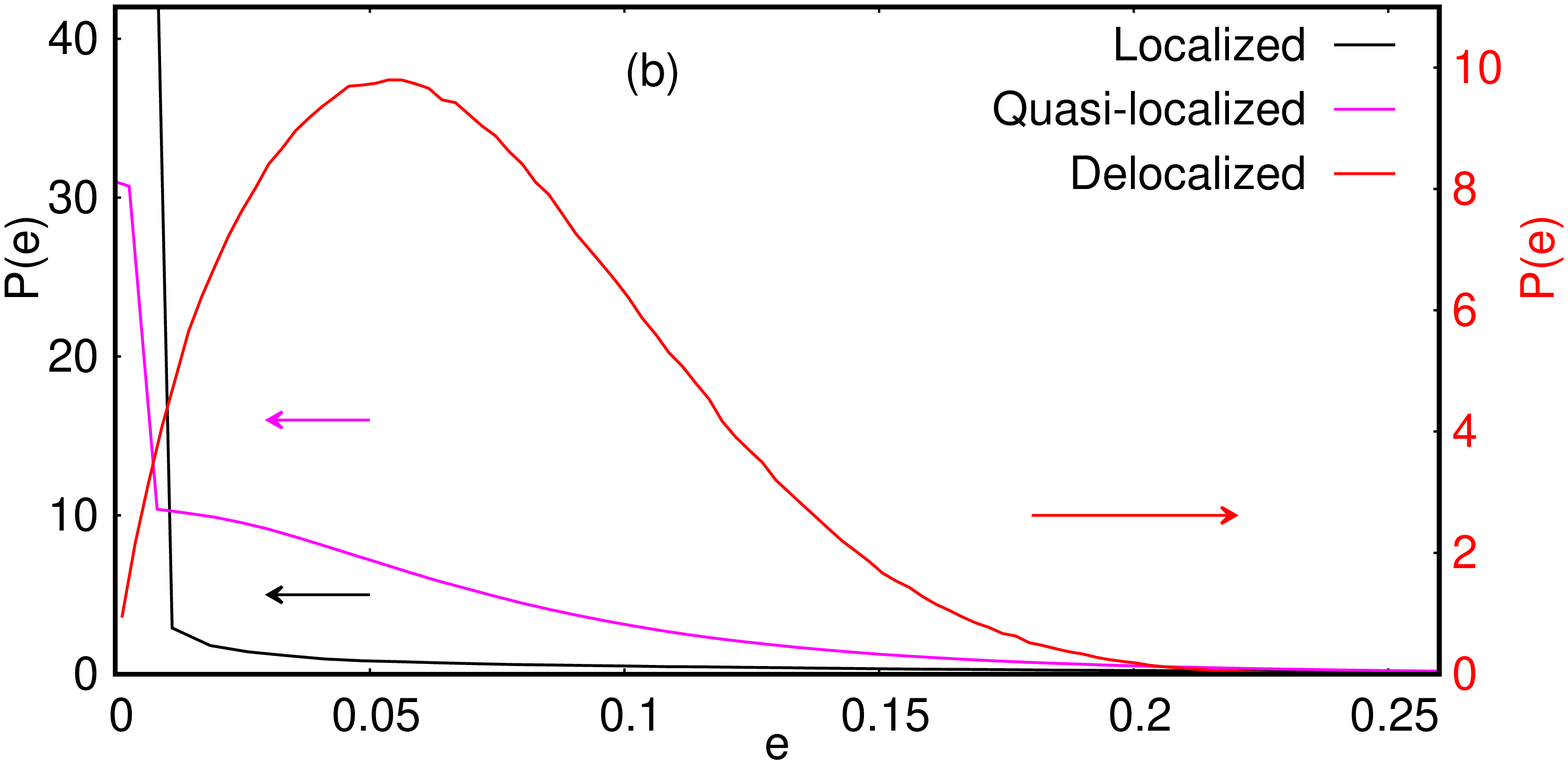}
\caption{ (Color online)(a) Distribution $P(s)$ of the spacings, $s$, of the eigenvalues corresponding to consecutive modes from three distinct windows of participation ratio (PR), at $T=0.050$. While dashed lines represent the true Poisson and the Wigner distributions, different types of points depict the level spacing distributions $P(s)$ for modes in a particular range of PR: For modes having PR$<0.05$ broadly follow Poisson distribution (localized modes), modes having PR$>0.35$ yield Wigner distribution (delocalized modes) and modes with $0.05<$PR$<0.35$ show deviation from both the Poisson and Wigner distributions lying roughly mid-way between the two, which we identify as the quasi-localized modes. Solid lines represent the appropriate Brody distributions associated with each type of mode with respective Brody parameter $k$ (see text for details). (b) Distribution $P(e)$ of the magnitude $e$ of the polarization vectors for all the modes in the same range of PR as in panel (a). For localized modes, the distribution is sharply peaked around zero and falls very rapidly ($P(e) \approx 120$ for $e \sim 0$ but we show $P(e)$ only up to $45$ for clarity). For delocalized modes $P(e)$ features a broad distribution, whereas, for the quasi-localized modes, there is a part where $P(e)$ is sharply peaked around zero and then there is a long tail.
}
\label{fig:IWM_QNM_LevelSpace}
\end{figure}

\subsection{Level spacing statistics}\label{Sec.LevelSpace}

In the past, the concepts of RMT played an instrumental role in discovering universality in a large variety of disordered systems, such as disordered mesoscopic systems, complex nuclei, quantum chaotic systems and even glass-forming systems~\cite{RevModPhys.72.895, PhysRevLett.54.1645, GUHR1998189, PhysRevE.64.016305, PhysRevLett.74.936}. Studies of the spectra of these systems, which are in general system dependent, have established that the statistical properties of the spectral fluctuations can be associated with one of the three universality classes identified in RMT~\cite{Mehta_RMT, GUHR1998189, RevModPhys.53.385, Matharoo}. Vibrational spectra, obtained from computer simulations, of several disordered solids and liquid systems have revealed that the spectral fluctuations follow the Gaussian orthogonal ensemble (GOE) of RMT~\cite{PhysRevE.64.016305, PhysRevLett.74.936, Matharoo,RevModPhys.53.385,PhysRevLett.81.136}. Particularly, these studies have shown that the delocalized modes conform to GOE statistics while the localized modes were found to obey Poissonian statistics~\cite{PhysRevLett.81.136,Clapa}. We study this issue here by evaluating the distribution of spacings between the successive eigenvalues of the Hessian matrix for different windows of participation ratios of the QNMs. Through this study, we also address the question of whether the suggested universality holds for trapped systems of long-range interacting particles. 

To identify the window of participation ratio within which modes can be quantified as localized, quasi-localized or delocalized, we compute level spacing distribution, $P(s)$, for various ranges of PR values. The level spacing distribution, $P(s)$, gives the probability of spacing $s$ between successive eigenvalues of the of the Hessian matrix. Thus, $s_i = (\lambda_{i+1}-\lambda_i)/\Delta$, where $\lambda_i$ is the $i$-th eigenvalue of the Hessian matrix (all eigenvalues are arranged in an ascending order) and $\Delta$ is the mean-level spacing.

Fig.~\ref{fig:IWM_QNM_LevelSpace}(a) shows $P(s)$ as a function of $s$ at $T=0.050$. We find that spacings between the eigenvalues follow a Poisson distribution for modes having PR$<0.05$ while it follows Wigner distribution for modes with PR$>0.35$. Thus, we can characterize the modes with PR$<0.05$ as localized modes while those having PR$>0.35$ as delocalized modes. The modes for which $0.05<$PR$<0.35$, $P(s)$ show significant deviation from both, the Poisson and the Wigner distributions and thus we identify these modes as quasi-localized as we discuss below.

To characterize the level spacing distribution, $P(s)$, further, we have fitted each distribution, as identified in Fig.~\ref{fig:IWM_QNM_LevelSpace}(a), using the Brody function~\cite{RevModPhys.53.385}
\begin{equation}
p_{k}(s)= (k+1)bs^k \mathrm{e}^{-bs^{k+1}}; b=\left[\Gamma \left(\frac{k+2}{k+1}\right)\right]^{k+1}
\end{equation}
In the definition, $k$ represents the nature of the distribution $P(s)$: $k= 0$ for Poisson distribution  while $k=1$ for Wigner distribution~\cite{RevModPhys.53.385}. We find a lower threshold of $k = 0.001\pm 0.0005$ for the localized modes while an upper threshold of $k = 0.85\pm 0.01$ (see Fig.~\ref{fig:IWM_QNM_LevelSpace}(a)) for the delocalized modes. For the quasi-localized modes, we find $k= 0.50\pm 0.03$ which lies at the middle of the two extreme values for $k$ representing Poisson and Wigner distributions. This validates our identification of the quasi-localized modes: something for which Brody parameter $k$, describing the nature of $P(s)$, is equally far away from the Poisson distribution and the Wigner distribution.

Though we justify our use of thresholds in PR at $0.35$ and $0.05$ to demarcate the boundary between delocalized and quasi-localized modes, and between quasi-localized and localized modes respectively, we admit that these cut-off values are chosen on some ad hoc basis. However, we assert that our conclusions remain unaltered if these cut-offs are changed within some margin. This is illustrated with an example in Appendix B.

\subsection{Distribution of the magnitude of polarization vectors}\label{sec.Emag}

One independent way to check whether the above separation of all the modes into localized, quasi-localized and delocalized is consistent or not, is to compute the distribution $P(e)$ of the magnitude, $e$, of the polarization vectors for the modes in each region mentioned above. Fig.~\ref{fig:IWM_QNM_LevelSpace}(b) shows that for localized modes, $P(e)$  is sharply peaked around zero implying that the contribution of most of the particles to the eigenvector in these modes is practically zero. Only a few particles have larger magnitude for the polarization vectors and thus contribute to the tail part of the distribution. On the other hand, for modes with PR$>0.35$, representing the delocalized part of the spectrum, $P(e)$ is peaked around a non-zero value and shows a broader distribution. Interestingly,  for the modes which are in between these two regimes, i.e. $0.05<$PR$<0.35$, $P(e)$ has two parts: a peak near zero (though the height of this peak is lower compared to that for localized modes) and then a long tail for finite values, implying that there are particles which have very small polarization vectors along with many particles with larger magnitude of the polarization vectors. Thus, quasi-localized modes have a localized and a delocalized part in the distribution $P(e)$.

\begin{figure}[t]
\includegraphics[width=8.5cm,keepaspectratio]{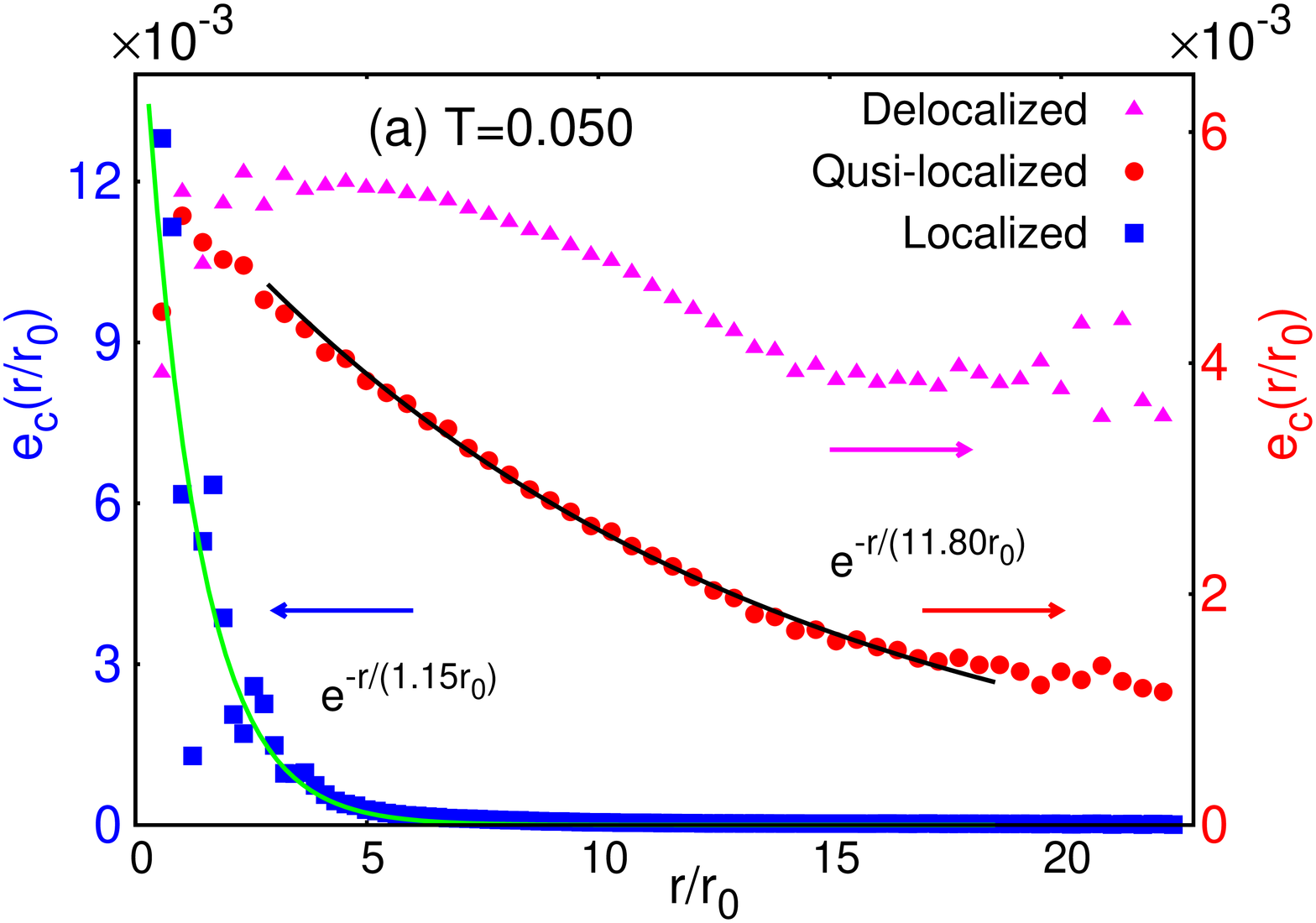}\\
\includegraphics[width=8.8cm,keepaspectratio]{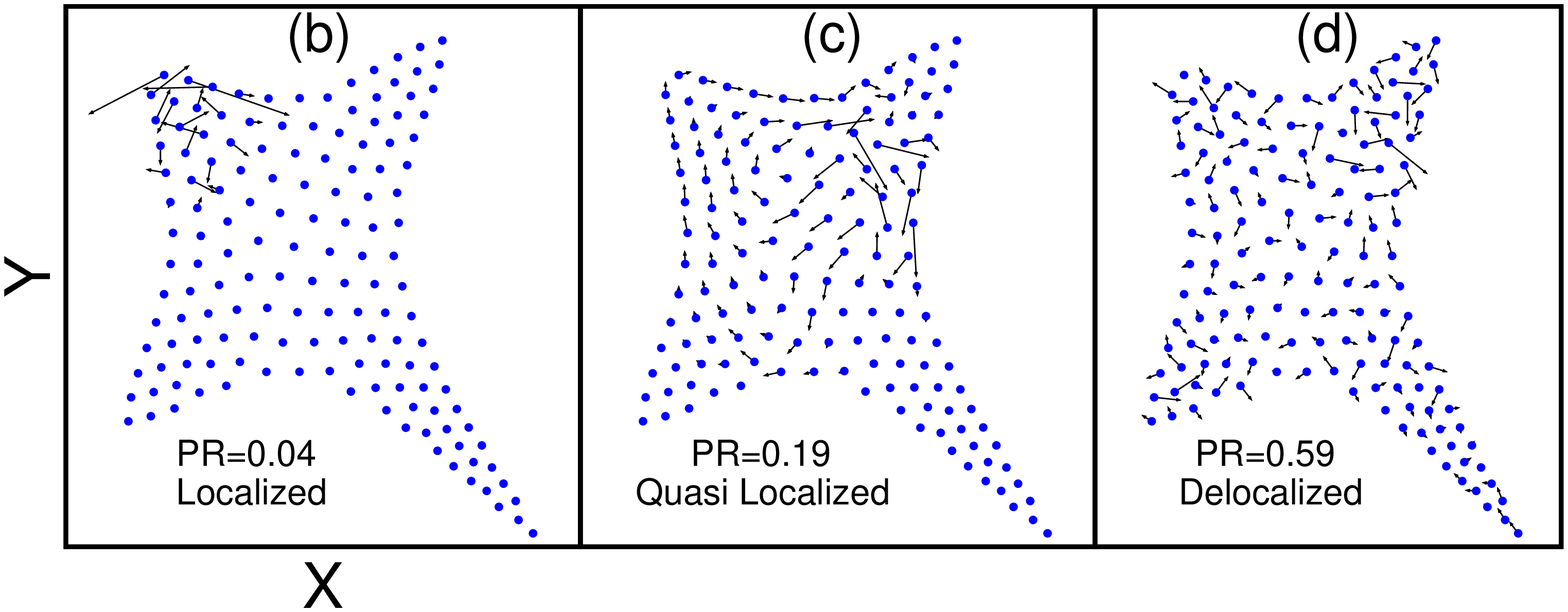}
\caption{ (Color online) The $r$-dependence of the correlation $e_c(r)$ between the magnitude of the polarization vectors of pair of particles a distance $r$ apart for the localized, quasi-localized and delocalized modes at $T=0.050$ are shown for a system with $N=150$ particles. Thick dots represent the actual data points while solid line is the exponential fit to the data points. Polarization vectors for $N=150$ particles in a typical (b) localized mode, (c) quasi-localized mode and (d) delocalized mode. The PR values for each mode is also mentioned in the plot. In panel (b-d) small dots represent the positions of the particles in the quenched configuration and the arrows depict the polarization vector for each particle. The length of each arrow is multiplied by a factor of $5$ for visual clarity.
}
\label{fig:IWM_QNM_Loclen}
\end{figure}

For the quasi-localized modes, particles with larger magnitude of the polarization vectors tend to form clusters in space. To quantify the average spatial extent of such clustered region, we compute the correlation between the magnitude of the polarization vectors of all the pair of particles. We define
\begin{equation}
e_c(\vec{r}) = \left\langle n(\vec{r}_i)n(\vec{r}_i+\vec{r}) \right\rangle
\label{Eq.Corrln}
\end{equation}
where, for a given mode $n(\vec{r}_i) = \vert \vec{e}(\vec{r_i}) \vert$ is the magnitude of the polarization vector of the $i$-th particle at the position $\vec{r}_i$. 

In Fig.~\ref{fig:IWM_QNM_Loclen}(a), we show the $r$-dependence of the correlation $e_c(r)$ for the localized, quasi-localized and delocalized modes as identified above. Here, distance $r$ between any pair of particles is expressed in units of average inter-particle distance, $r_0$. We find that for localized and quasi-localized modes $e_c(r)$ decays exponentially with $r$: $e_c(r) \propto \exp[-r/\xi]$. While for localized modes $e_c(r)$ decays very rapidly, it falls slowly for quasi-localized modes. On the other hand, for delocalized modes $e_c(r)$ shows a very weak $r$-dependence; it remains almost flat. By fitting the individual curves for localized and quasi-localized modes, we find that $\xi_{\mathrm{loc}} \sim r_0$ while $\xi_{\mathrm{qloc}} \sim 12 r_0$. We find that the $r$-dependence of $e_c(r)$ for different types of modes is almost independent of $T$. Thus, our study provides quantitative identification of the quasi-localized modes in disordered and long-range interacting systems. In Appendix B, we further demonstrate that the qualitative features of the distribution $P(e)$ and correlation $e_c(r)$ remain unchanged upon small tweaking of the boundary between the quasi-localized and delocalized modes.

In Fig.~\ref{fig:IWM_QNM_Loclen}(b-d), we show the polarization vectors for $N=150$ particles for a typical localized mode (panel b), quasi-localized mode (panel c) and delocalized mode (panel d). We can see that for the localized mode, only a few particles contribute to the mode while for the delocalized one, almost all the particles contribute to some extent. For the quasi-localized mode, there are several particles which have relatively large magnitude for the polarization vector and the rest of the particles have small but nonzero contributions to the mode.

In Fig.~\ref{fig:IWM_QNM_Loclen}(c), we see that the particles which contribute to the quasi-localized modes, due to larger magnitude of their polarization vectors, appear to be spatially  clustered. Thus, we naturally ask: What is the typical size of such a cluster, and how does that corroborate with the length scale $\xi_{\mathrm{qloc}}$ estimated from $e_c(r)$?

To define a cluster, based on the magnitude of the polarization vectors for the quasi-localized modes, we need two parameters: (a) a cut-off, $e_t$, for the magnitude of the polarization vectors, $e$, such that particles having $e \ge e_t$ are eligible to be part of a cluster, and (b) a cut-off distance, $r_c$, of a particle from a cluster to decide whether it also belongs to that particular cluster. With these two inputs we can group the particles in disjoint clusters of different sizes, $n$. A cluster of size $n$ implies that there are $n$ particles which belong to that cluster and thus, for 2D systems, the typical length scale associated with such a cluster can be considered as $\sqrt{n}$. In our analysis, $r_c$ is chosen as the position of the first minimum of the pair correlation function.
\begin{figure}[t]
\includegraphics[width=8.5cm,keepaspectratio]{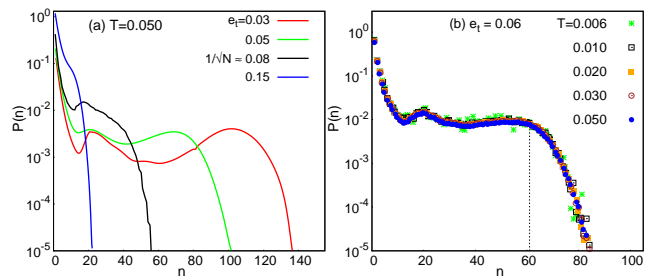}
\caption{ (Color online) (a) Probability  distribution $P(n)$ of cluster size $n$ for different values of the cut-off, $e_t$, for the magnitude of the polarization of vector, $e$, for the quasi-localized modes at $T = 0.050$ in a semi-logarithmic plot. $N=150$ is the total number of particles.
(b) Probability  distribution $P(n)$ of cluster size $n$ for various $T$ in a semi-logarithmic plot with $e_t = 0.06$. The dashed vertical line indicates the value of $n (=61)$ beyond which $P(n)$ decreases monotonically.
}
\label{fig:Cluster}
\end{figure}

The immediate question arises: What would be an appropriate choice for $e_t$? In order to address this, we present the probability distribution, $P(n)$, of cluster size $n$, for a range of choices of $e_t$ in Fig.~\ref{fig:Cluster}(a). We see that for small magnitude of $e_t$, say, $e_t=0.03$, $P(n)$ exhibits a second peak at a macroscopic value of $n\sim 110$, in addition to a stronger peak for smallest $n$. With the increase of $e_t$, the height of the peak at large $n$ decreases and the peak moves down to smaller values of $n$ as well, finally making $P(n)$ to feature only a monotonic decay for $e_t \gtrsim 0.08$. We also note here that $P(n)$ features an intriguing hump at intermediate values of $n$ for a range of $e_t (\leq 0.08)$. Its origin is under investigation, and we do not have a clear understanding. Let us now consider $e_t=0.06$, a value which is about $10\%$ of the average inter-particle distance -- a typical threshold used to determine melting by (diffusive) delocalization in Lindemann's description~\cite{Lindemann,DA13}. For this value of $e_t=0.06$, $P(n)$ develops for the first time a tendency towards formation of a large cluster, as signalled by the peak at large $n$, as seen from Fig.~\ref{fig:Cluster}(b). This figure also illustrates that $P(n)$ is fairly insensitive to $T$ - it remains unaltered as the temperature is changed over a fairly large range. Thus, for $e_t=0.06$, it is more probable to find a typical cluster size $n\sim 61$, which spans over a large part of the system (note that $N=150$ in our analysis), or not having any cluster at all (we do not qualify a cluster of size $n=1$ as a `cluster'). Such a scenario is similar to the one shown in Fig.~\ref{fig:IWM_QNM_Loclen}(c). Thus, the typical length scale associated with the larger cluster becomes $\sqrt{61} r_0 \sim 8 r_0$ which is in good agreement with the value $\xi_{\mathrm{qloc}} \sim 12 r_0$ we found from $e_c(r)$.

Are there additional physical inputs to support the above choice of $e_t=0.06$ as a reasonable cut-off? We note that the above choice is very close to $N^{-1/2} \sim 0.08$; $N=150$ being the total number of particles in the system. Such a choice is motivated by the fact that if all the $N$ particles participate in a mode equally (representing a perfectly delocalized mode) then each particle would have the magnitude of polarization vectors $N^{-1/2}$. Our choice of $e_t=0.06$ is close to $N^{-1/2}$ as well, and appears to be a reasonable cut-off based on above observations.

Having argued for our identification of optimal $e_t$, we must also admit that a small variation around this value changes the nature of $P(n)$ smoothly and continuously. Fig.~\ref{fig:Cluster}(a) shows $P(n)$ versus $n$ for several values of $e_t \in [0.03, N^{-1/2}]$ at $T=0.050$. The continuous evolution of this distribution raises uncertainty in the robustness of the cluster size of the quasi-localized modes. As discussed in Appendix B, the form of the cluster size distribution and its dependence on the choice of $e_t$ remain qualitatively unchanged as the values of the cutoffs used to define quasi-localized modes are changed within a limited range.

Thus, by analyzing the level spacing statistics and the correlation function $e_c(r)$, we can divide the whole quenched normal mode spectrum into three sections: localized, quasi-localized and delocalized modes. Now, the questions is: What is the role of these modes in dictating the long time dynamics of the particles? Can we identify the particles which exhibit large displacements at long time looking at the normal modes associated with the initial configuration? A striking correlation between the regions of motion in the low frequency modes and the regions of high mobility~\cite{PhysRevLett.96.185701} has been postulated in both two- and three-dimensional supercooled liquids. We set out to examine such correlations in our systems.

\section{Correlating normal modes with displacements of mobile particles}\label{Sec:dispcorln}

In order to study the time $(\Delta t)$ dependence of the correlation of displacement and normal modes at a given $T$, we have defined two generalized lists $C_d$ and $C_e$ each of length $N$ where $N$ is the total number of particles in the system. For a given time interval $\Delta t$, $C_d(i,t_0 + \Delta t) = 1$ if $i$ represents a `fast' particle in that interval; otherwise $C_d(i, t_0 + \Delta t) = 0$. At any given $\Delta t$, we consider top $20\%$ of the particles with the largest magnitude of displacement as fast particles. On the other hand, $C_e(i,t_0)$ contains information of the magnitude of the polarization vectors $|e_l^i|$ in certain $N_e$ number of quenched normal modes, obtained for the IS corresponding to the configuration at the initial instant $t_0$. More explicitly, we calculate $E_i= \frac{1}{N_e} \sum_{l=1}^{N_e} |e_l^i|^2$, and define $C_e(i,t_0)=1$ for those $i$-th particles for which $E_i$ lies in the top $20\%$ bracket of its value.
Now, we can define a time dependent correlation function
\begin{equation}
C_N(\Delta t) = \frac{\sum_{i=1}^{N}\langle C_d(i,\Delta t)C_e(i,t_0) \rangle}{N_f}
\end{equation}
where $\langle . \rangle$ represents an average over independent time origins, $t_0$ and $N_f$ is the number of particles considered to be fast (in this case top $20\%$ of $N$, having largest displacements). From the definition $C_N(\Delta t) =1$ only when both the lists have the same particles implying maximum correlation, and is zero when the two lists have no common particle.
\begin{figure}[t]
\includegraphics[width=8.5cm,keepaspectratio]{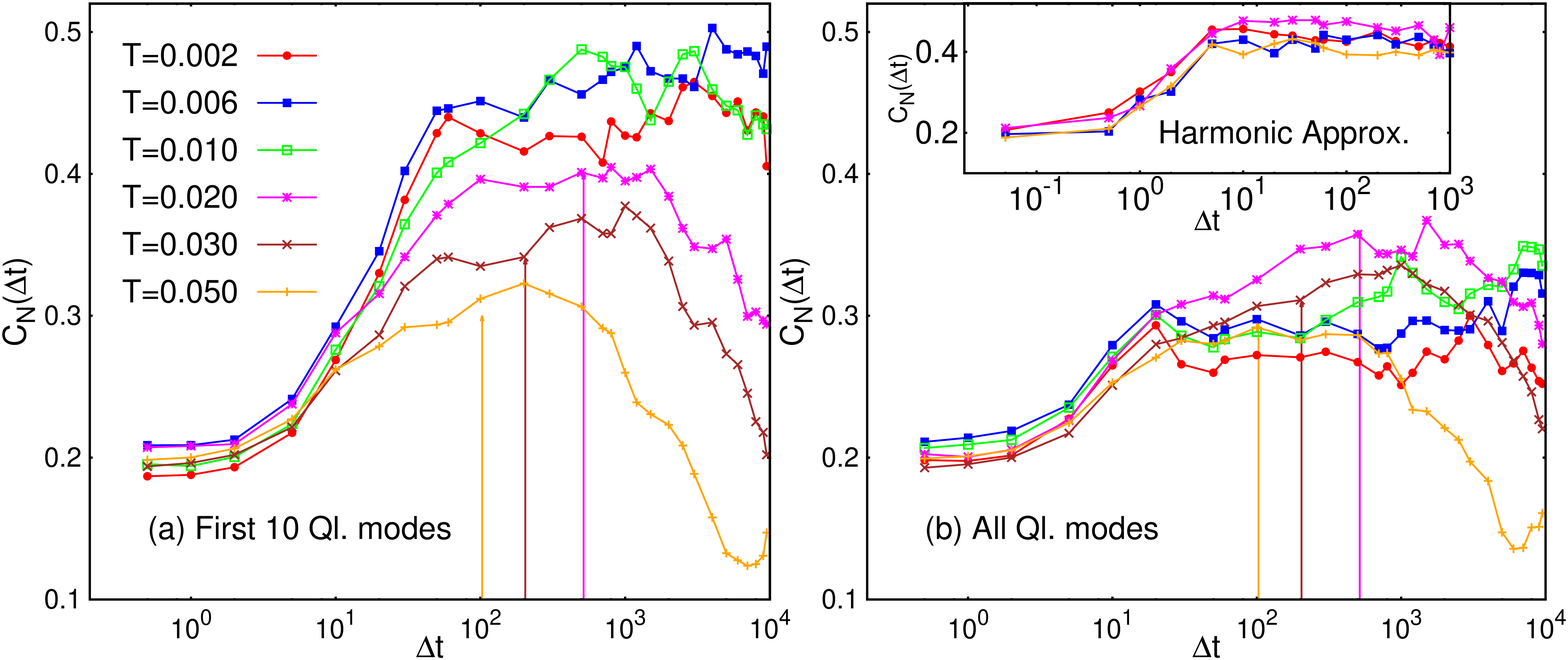}\\
\hspace{0.1cm}\includegraphics[width=4.2cm,keepaspectratio]{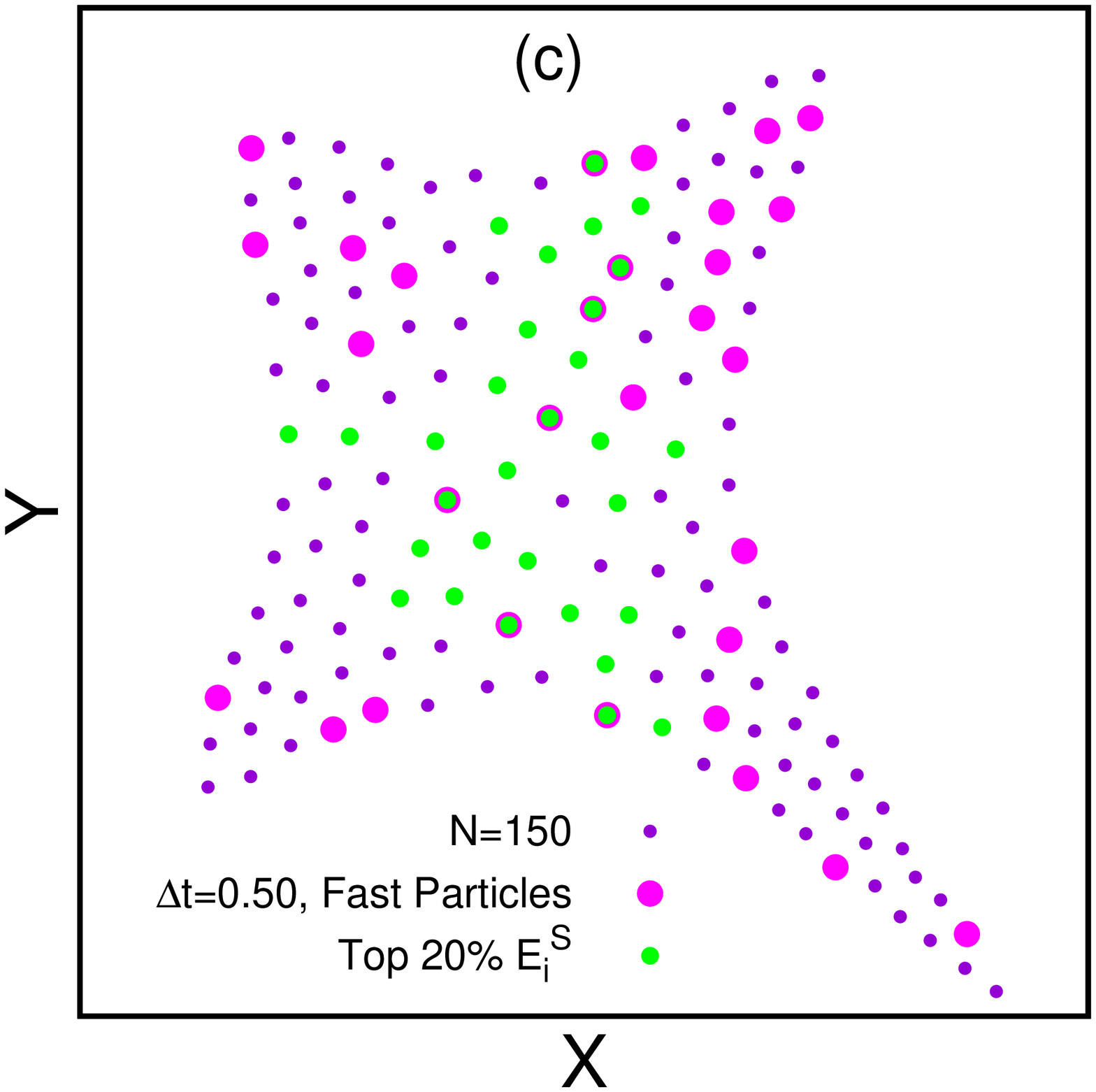}
\includegraphics[width=4.2cm,keepaspectratio]{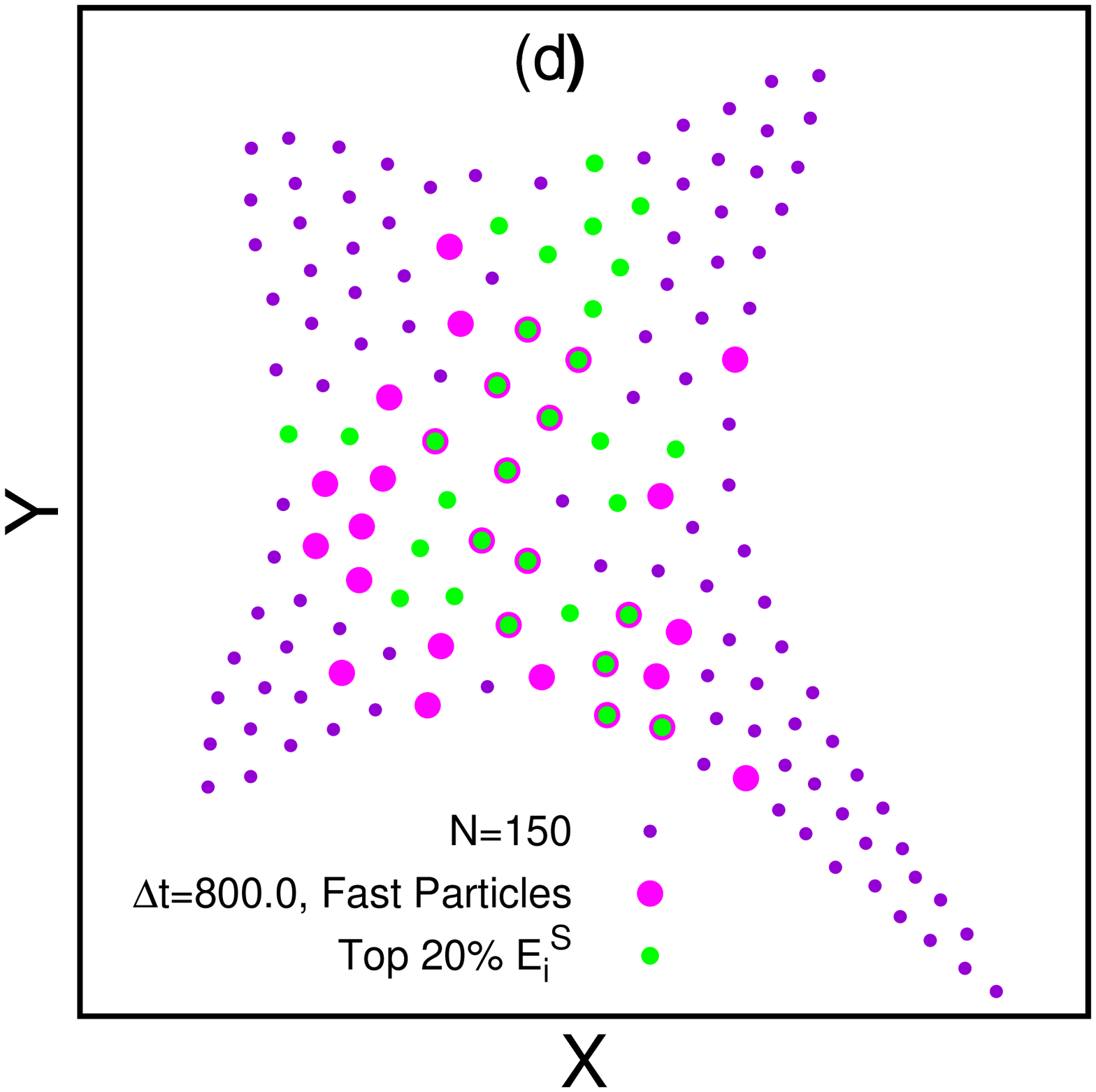}
\caption{ (Color online) The time dependence of the correlation $C_N(\Delta t)$ between the fast particles and particles having larger values of $E_i$ (i.e. for the sum of the magnitude of the polarization vectors over the certain quasi-localized modes, see text) for different $T$. $N_e$ is the number of eigen modes considered to compute $E^s_i$. (a) Only first ten $(N_e=10)$ quasi-localized modes are considered. (b) All the low-frequency ($\omega \lesssim 1$) quasi-localized modes are considered. Inset of panel (b): The time dependence of the correlation $C_N(\Delta t)$ for the same parameters as in main panel (a) but displacement of a particle is computed under harmonic approximation. Arrows in panel (a) and (b) represent the value of $C_N(\Delta t)$ at $\Delta t=\tau_{\alpha}$, the structural relaxation time, for a given $T$. Snapshot showing the typical correlation $C_N(\Delta t)$ for (c) $\Delta t = 0.50$ and (d) $\Delta t = 800.0$ for $T=0.006$.
}
\label{fig:IWM_QNM_Disp_Corrln}
\end{figure}

Fig.~\ref{fig:IWM_QNM_Disp_Corrln}(a) shows the time dependence of the correlation $C_N(\Delta t)$ for different $T$. Here, we consider $N_e=10$, only the first $10$ \textit{quasi-localized} modes. For all $T$, we find that $C_N(\Delta t)$ is smaller for $\Delta t <5$ and then the correlation starts to increase. For low temperature $(T \leq 0.006)$, we see a good correlation between the fast particles and those with large polarization vectors in the initial low-energy quasi-localized modes for long time intervals as $C_N(\Delta t)$ tends to saturate; it fluctuates around a mean value at long time intervals. But, at high temperatures $(T>0.010)$, $C_N(\Delta t)$ decays after attaining a maximum value at some intermediate time which decreases with increasing $T$. While we see that the correlation between the low-energy quasi-localized modes and fast particles is not perfect -- it reaches only up to $0.5$, the presence of such correlation is statistically profound and is present for all temperatures. In fact, persistence of such correlation even at temperatures as large as $T \sim 0.050$, which is more than two times the crossover temperature $T_X (\sim 0.020)$, is very intriguing.

The reason for the final saturation of the $C_N(\Delta t)$ at low $T$ can be attributed to the fact that the system explores very few distinct inherent structures because of lower thermal energy; it remains in the basin of a favorable inherent structure for a long time. At short times, particles execute random vibrational motion around their equilibrium positions and that random motion cannot be described by only a few low-energy quasi-localized modes; all the eigen modes are necessary for representing such uncorrelated motion. Consequently, for a given temperature, $C_N(\Delta t)$ attains a small value for small $\Delta t$. But at long time, we see that it is the initial quasi-localized modes that dictate the displacements of the fast particles. Thus, we see that the initial low lying modes are capable of identifying particles which will undergo large displacements at long time interval. 

Fig.~\ref{fig:IWM_QNM_Disp_Corrln}(b) shows the time dependence of the correlation $C_N(\Delta t)$ for different $T$ when all the low-frequency ($\omega \lesssim 1$) quasi-localized modes are considered. While the time dependence shows qualitatively similar behavior, the value of the correlation becomes smaller at any given time interval. The qualitative similarity between Fig.~\ref{fig:IWM_QNM_Disp_Corrln}(a) and Fig.~\ref{fig:IWM_QNM_Disp_Corrln}(b) hints that it is only the low-energy quasi-localized modes that are responsible for the heterogeneous dynamics at long times in the system.

The smaller values of $C_N(\Delta t)$ at small $\Delta t$ and its saturation at low $T$ and large $\Delta t$ can also be explored from the harmonic approximation~\cite{ziman}, which is considered to be a good approximation for describing the very low $T$ behavior of solids. Under this approximation, each particle executes small amplitude vibrational motion where the typical amplitude is proportional to $\sqrt{T}$. To compute $C_N(\Delta t)$ under harmonic approximation, we first obtained the configurations of the particles for each $T$ using the method described in Appendix~\ref{Ap:HM} (See Eq.~\ref{Eq:Harmnc_Approx}) and then followed the same steps as described above. Inset of Fig.~\ref{fig:IWM_QNM_Disp_Corrln}(b) shows $C_N(\Delta t)$ at different $T$ where displacement of the particles are computed under the Harmonic approximation. Even in this case, we see that $C_N(\Delta t)$ is small for small $\Delta t$ and shows saturation at long time for all $T$. Within the harmonic approximation, we do not expect $C_N(\Delta t)$ to decrease at long times as particles are always confined inside the basin of an IS and thus the dynamics remains correlated with the initial eigen modes. Fig.~\ref{fig:IWM_QNM_Disp_Corrln}(c,d) show snapshots depicting the typical correlation $C_N(\Delta t)$ for small $(\Delta t = 0.50)$ and large $(\Delta t = 800.0)$ time intervals.

We would like to emphasize that the importance of low-frequency quasi-localized modes in describing the heterogeneous dynamics at long time has already been discussed in the context of supercooled liquids~\cite{Widmer_Natphys,Widmer_JCP, Brito07, PhysRevLett.105.025501,PhysRevLett.104.248305,PhysRevLett.107.188303}. But in this work, by introducing the correlation function $C_N(\Delta t)$, we have quantified the importance of such low frequency modes in describing the long time dynamics even in the case of finite systems of Coulomb interacting particles. Further, we establish that it is not just the low frequency modes but the low frequency quasi-localized modes which are responsible for the observed heterogeneous dynamics in Coulomb clusters. For this reason, independent identification of the quasi-localized modes is necessary which we developed in sec.~\ref{sec:VDOS}.

In Fig.~\ref{fig:IWM_QNM_Disp_Corrln}(a-b), we see that $C_N(\Delta t)$ attains a maximum value at some intermediate time which shifts towards lower values with increasing $T$. What does this time scale signify? Is it related to the structural relaxation time, $\tau_{\alpha}$, of the system? Note that the distinction between `slow' and `fast' particles is most relevant around $\tau_{\alpha}$. So, we, next, compute the structural relaxation time, $\tau_{\alpha}$, at different $T$ for the system.

\begin{figure}[t]
\includegraphics[width=8.5cm,keepaspectratio]{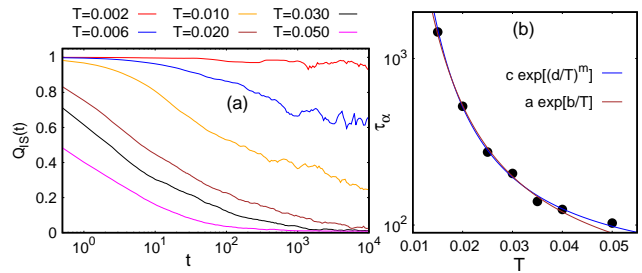}
\caption{ (Color online) (a) The $t$-dependence of the overlap function $Q_{IS}(t)$ computed considering the \textit{quenched configurations} for different $T$. We find that at small $T$, it decays very slowly while for $T \geq 0.02$, $Q_{IS}(t)$ decays to zero at long time. (b) The $T$-dependence of average structural relaxation time, $\tau_{\alpha}$. It is obtained from the area under the $Q_{IS}(t)$ traces for temperatures where $Q_{IS}(t)$ decays to zero at long time. Solid lines represent the best fit to the numerical data and shown for two different functional forms (See text for details). The best fit parameters are: $c=48.1, d=0.04, m=1.31$ and $a=27.6, b=0.06$.
}
\label{fig:IWM_QIS}
\end{figure}

For bulk systems, $\tau_{\alpha}$ is generally estimated from the long-time behavior of the self part of the intermediate scattering function $F_s(k,t)$~\cite{SmarajitRev1,Sastry1}, where the value of $k$ is usually taken as the wavenumber for which the static structure factor exhibits the first peak. Since, for small finite systems, such as our irregular confinement, description in terms of quantities in reciprocal space ($k$-space) is not a natural choice~\cite{PRE17}, we use the overlap function~\cite{SmarajitRev1} which is defined in terms of position space coordinates. While $\tau_{\alpha}$ can also be estimated using the equilibrium MD configurations, an estimate using the quenched configurations has the advantage that it helps to get rid of the contribution of small amplitude vibrational motion in the overlap function and thus bring out the true long time dynamical behavior of the system which is free of spurious effects. Thus, to estimate $\tau_{\alpha}$, we compute the temperature dependence of the overlap function, $Q_{IS}(t)$, considering the quenched configurations only. Denoting $\vec{r}_{IS}^{~i}(t)$ as the position of the $i$-th particle in the IS corresponding to the equilibrium configuration at time $t$, we define $Q_{IS}(t)$ as~\cite{SmarajitRev1}: 
\begin{equation}
Q_{IS}(t) = \left\langle \frac{1}{N} \sum_{i=1}^{N} w(|\vec{r}_{IS}^{~i}(t_0 + t) - \vec{r}_{IS}^{~i}(t_0)|)\right\rangle
\end{equation}
where, $w(r) = 1.0$ if $r < r_c$ and zero otherwise. The angular parentheses denotes averaging of results over the time origin, $t_0$, and also over different realizations of the disorder. We choose $r_c=0.15r_0$. Fig.~\ref{fig:IWM_QIS}(a) shows the $t$-dependence of $Q_{IS}(t)$ for different $T$. We find that at small $T$, it decays very slowly while for $T \geq 0.015$, $Q_{IS}(t)$ decays to zero at long time.

We estimate $\tau_{\alpha}$ from the area under the trace of $Q_{IS}(t)$ versus $t$ for temperatures where $Q_{IS}(t)$ decays to zero at long time. The value of $\tau_{\alpha}$ thus obtained for different $T$ is shown in Fig.~\ref{fig:IWM_QIS}(b) (also in Fig.~\ref{fig:IWM_QNM_Disp_Corrln}(a-b) through the starting position of the vertical arrows).  
So, we see that the time at which $C_N(\Delta t)$ attains maximum value at high $T$ is comparable to the structural relaxation time at that $T$ (Fig.~\ref{fig:IWM_QNM_Disp_Corrln}(a-b)). This is also consistent with the physical expectation that the system remains in the basin of a low-energy inherent structure for a duration of the order of structural relaxation time.

We also find that $\tau_{\alpha}$ increases quite rapidly with decreasing $T$ (Fig.~\ref{fig:IWM_QIS}(b)). To understand the $T$-dependence of $\tau_{\alpha}$, we fit the data in Fig.~\ref{fig:IWM_QIS}(b) with several functional forms and the best fit has been identified using the $\chi^2$ analysis. Given the small number of data points as well as the quality of the statistics, our estimates cannot choose between an Arrhenius behavior$(\tau_{\alpha}(T)=a \exp[b/T]$ with $a=27.6$ and $b=0.06)$ and an Avramov-Milchev ($\tau_{\alpha}(T)=c \exp[\left( d/T \right)^m ]$ with $c=48.1, d=0.04,$ and $m=1.31$)~\cite{AVRAMOV} type behavior. We find that these two functional forms capture the $T$ dependence of $\tau_{\alpha}$ rather accurately for lower temperatures. We also confirm that our results do not conform to the Vogel-Fulcher-Tammann (VFT) form~\cite{SmarajitRev1} or a power-law behavior~\cite{Sastry1} with a similar degree of accuracy.

So far we have analyzed the dynamics of particles in irregular confinement with respect to the quenched normal modes. We can also evaluate the Hessian matrix using the instantaneous equilibrium configurations and normal modes obtained in this way are called the instantaneous normal modes (INMs). Below, we discuss some of the features of the INMs.

\section{Instantaneous normal modes}\label{INM}

The instantaneous normal modes, which play important role in understanding the solid and liquid states~\cite{PhysRevLett.74.936,INM_d2,PhysRevLett.108.225001}, are extension of the conventional harmonic normal-mode approach. The INM spectrum carries information about the average curvature of the instantaneous (equilibrium) potential energy landscape. 
\begin{figure}[t]
\includegraphics[width=8.5cm,keepaspectratio]{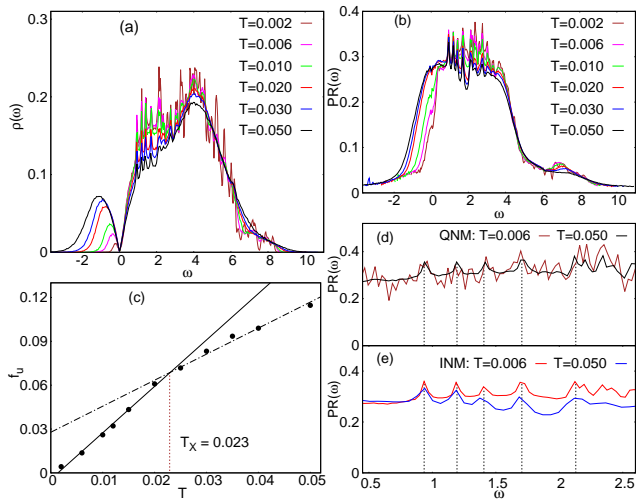}
\caption{ (Color online) (a) Density of states of instantaneous normal modes as a function of frequency are shown for $N=150$ particles in irregular confinement at different $T$. The unstable modes (imaginary frequencies) are shown on the negative frequency axis. (b) The average participation ratio PR$(\omega)$ as a function  of $\omega$ for the various spectra shown in (a). Results are enhanced by obtaining statistics over $6$ independent realizations of irregularity parameters. (c) $T$-dependence of the fraction of unstable modes $(f_u)$ obtained from INMs. Thick lines are the linear fit to the actual data points. Crossing of these two lines gives an estimate for the cross-over temperature $T_X (\sim 0.023)$ for the system. PR for (d) QNM and (e) INM at low $T (=0.006)$ and high $T (=0.050)$ for $\omega<3.5$. The peaks which persist for all $T$ (indicated by vertical dashed lines), appear at the same $\omega$ values for INM and QNM. 
}
\label{fig:IWM_INM}
\end{figure}
Since at any finite temperature, an arbitrary configuration may not be at the potential minimum, the associated Hessian matrix, in general, will not be positive definite and thus can contain negative eigenvalues. Therefore, We can categorize the INMs as stable (positive eigenvalues or real frequencies) and `unstable' (negative eigenvalues or imaginary frequencies) modes. We obtain the normalized INM density of states, $\rho(\omega)$, by averaging over many independent configurations at a given $T$. Fig.~\ref{fig:IWM_INM}(a) shows the density of states, $\rho(\omega)$, for INM as a function of frequency at different temperatures. For the lowest $T (=0.002)$, $\rho(\omega)$ shows a very peaked structure, implying that only few modes at particular frequencies can exist in the ordered state. There are also a few unstable modes, shown on the negative frequency axis. 

With increasing temperature, relatively continuous mode spectrum is observed, indicating the disordered arrangement of the particles in the system. It is interesting to note that there are certain modes (at the low $\omega$ region) which are quite robust to $T$. Fig.~\ref{fig:IWM_INM}(b) shows average participation ratio (PR) as a function of mode frequency, $\omega$, for various spectra shown in Fig.~\ref{fig:IWM_INM}(a). We find that PR is higher for intermediate modes in the spectrum and lower for both low and high-frequency modes. This is similar to what we observed for QNMs. 

The unstable part of the density of states (as shown on the negative frequency axis), indicating the liquid-like behavior, becomes wider with increasing $T$. Thus, the fraction of unstable modes $(f_u)$ increases, compared to the total number of modes, as $T$ goes from $0.002$ to $0.050$. These results qualitatively reflect the thermal evolution of our system from a solid-like to a liquid-like state. Such an identification of solid to liquid transition in terms of INMs is recently studied experimentally for charged particles in harmonic (parabolic) trap~\cite{PhysRevLett.108.225001}.

We show the $T$-dependence of the fraction of unstable modes, $f_u$, in Fig.~\ref{fig:IWM_INM}(c). One can estimate the crossover temperature $T_X$ from the $T$-dependence of $f_u$. There is a change in slope around $T \sim 0.023$ which we identify as $T_X$. The value of $T_X$ found this way remains very close to the value obtained from earlier studies $(T_X =0.02)$~\cite{PRE17,DA13}. For INM, while there are few low frequency modes which are robust to $T$, interestingly, we find that such robust peaks appear around the same $\omega$ values for both INM and QNM. Fig.~\ref{fig:IWM_INM}(d-e) show such similarities in the appearance of the robust peaks in terms of PR at two selected temperatures for QNM and INM.

\section{Conclusion}\label{sec:conclude}   

In summary, we have computed the normal mode spectrum for Coulomb interacting particles in irregular confinement. We have classified the full quenched normal mode spectrum, based on participation ratio and the concepts of random matrix theory, in three groups: (a) localized, where the number of particles contributing to the mode is intensive, (b) delocalized, implying that the number of particles taking part in the mode is extensive, and (c) quasi-localized which is in between the localized and delocalized parts of the spectrum. Our analysis shows that the low frequency quasi-localized modes of the Hessian matrix of the IS corresponding to the initial configuration have good correlation with the dynamics of the particles over a time scale of the order of the structural relaxation time. In particular, we show that the particles with larger contribution to the sum of the squares of the polarization vectors of a small subset of low-frequency quasi-localized modes associated with the initial configuration are more likely to experience longer displacement at later times. Thus, we have identified the characteristic feature of a given configuration that gives rise to the heterogeneous dynamics in Coulomb clusters. From the analysis of the instantaneous normal modes, we estimate the crossover temperature which is close to what was reported from the analysis of static and dynamic properties of the same system~\cite{PRE17,DA13}. In recent studies on instantaneous normal modes, the main focus was the fraction of unstable modes which are closely associated with the self-diffusion constant~\cite{PhysRevLett.74.936,INM_d2} of the system. Thus, it would be interesting to study the temperature dependence of the diffusion constant for our system from the perspective of the instantaneous normal modes.

\vspace*{-0.4cm}
\section*{ACKNOWLEDGEMENTS}

BA thank Pranab Jyoti Bhuyan for valuable discussions. We acknowledge computational facilities at IISER Kolkata and IISc Bangalore. BA acknowledges University Grant Commission (UGC), India, for doctoral fellowship. AG and CD acknowledge the hospitality of the International Centre for Theoretical Sciences (ICTS) during a workshop (Code: ICTS/ispcm/2018/02), where part of this research was carried out.

\appendix

\section{Details of the harmonic approximation}\label{Ap:HM}

Here, we shall derive the equations used to compute the position of the particles under the harmonic approximation. Within this framework, we write the total potential (potential due to inter-particle interactions and also from the confinement) $V(\vec{r}_1, \vec{r}_2, \cdots, \vec{r}_N) \equiv V(\{ \vec{r}_i\})$, Taylor expanded up to the second order about a local minimum configuration $\{\vec{r}_i^{~0}\}$, as~\cite{ziman}:
\begin{align}
V(\{ \vec{r}_i\}) &\simeq V(\{\vec{r}_i^{~0}\}) + \frac{1}{2}\sum_{i,j=1}^{N}\sum_{\alpha,\beta=1}^{2} K_{i,j}^{\alpha,\beta} u_i^{\alpha}u_j^{\beta}
\label{Eq:Pot_Hm_Approx}
\end{align}
where,
\begin{align}
K_{i,j}^{\alpha,\beta} = \left[ \frac{\partial^2 V}{\partial r_{i}^{\alpha}\partial r_j^\beta} \right]_{\{\vec{r}_i^{~0}\}}
\label{Eq:Pot_Hm_Approx2}
\end{align}
Here, $u_i^{\alpha}$ is the $\alpha (=1,2 \equiv (x,y))$ component of the displacement of the $i$-th particle from its position $\vec{r}_i^{~0}$ at the local minimum.
Thus, the equation of motion for the $i$-th particle becomes:
\begin{equation}
\ddot{u}_i^{\alpha} = - \sum_{j=1}^{N} \sum_{\beta=1}^{2} K_{i,j}^{\alpha,\beta} u_j^{\beta}
\label{Eq:HM_EOM1}
\end{equation}
All particles are assumed to have equal mass which is set to unity. Let us make the following change of variables: $u_i^{\alpha} \rightarrow q_a = q_{2(i-1) + \alpha}$ where $a=1,2,\cdots 2N$, so that Eq.~\ref{Eq:HM_EOM1} becomes
\begin{equation}
\ddot{\textbf{q}} = -\textbf{\underline{K}} \textbf{q}
\label{Eq:HM_EOM4}
\end{equation}
where $\textbf{q}$ is a $2N \times 1$ column matrix and $\textbf{\underline{K}}$
is a $2N \times 2N$ matrix. Now, applying an orthogonal transformation matrix $\textbf{\underline{S}}$, we diagonalize $\textbf{\underline{K}}$ in a new basis, say,
$\textbf{h}$, so that: $\textbf{h} = \textbf{\underline{S}} \textbf{q}$, and we finally obtain the equations of motion in $\textbf{h}$-basis as,
\begin{equation}
\ddot{\textbf{h}} = -\textbf{\underline{L}} \textbf{h}
\label{Eq:HM_EOM5}
\end{equation}
Here, $\textbf{\underline{L}} = \textbf{\underline{S}}~\textbf{\underline{K}}~\textbf{\underline{S}}^{-1}$ is a diagonal matrix. This diagonalization of $\textbf{\underline{K}}$, the force-constant matrix (also called the Hessian matrix), yields the normal modes as eigenvectors $e_{a}^{n}$, and corresponding squared normal-mode frequencies as eigenvalues $\lambda_n = \omega_n^2$. Here $n$ is the normal mode eigen-index and $a$ represents the components of the eigenvectors. Thus, the solution of the Eq.~\ref{Eq:HM_EOM5} is
\begin{equation}
h_a(t) = h_a(0) \cos \omega_a t + \frac{\dot{h}_a(0)}{\omega_a} \sin \omega_a t
\end{equation}

We can now use the inverse transformations $(\textbf{h} \rightarrow \textbf{q} \rightarrow \textbf{u})$ in order to obtain the trajectories of the particles,
\begin{align}
r_i^{\alpha}(t) &= r_i^{0,\alpha} + \nonumber \\ 
                & \sum_{b=1}^{2N} e_{2(i-1) +\alpha}^{b} \left[ h_b(0) \cos \omega_b t + \frac{\dot{h}_b(0)}{\omega_b} \sin \omega_b t \right]
\label{Eq:Harmnc_Approx}
\end{align}
Thus, to generate the configurations of $N$ particles at any time $t$ under harmonic approximation, we need to choose these initial coordinates $r_i^{0,\alpha}$ appropriately. We have to consider a configuration of the system which minimizes the potential energy and thus satisfy the requirements for harmonic theory. From any given equilibrium MD configuration, we can generate such an energy minimized configuration (or inherent structure) for the system using conjugate gradient method~\cite{Num_Recipe}. To make sure that the obtained configuration is a stable one, the lowest eigenvalue of the dynamical matrix is be checked to be positive.

\begin{figure}[t]
\includegraphics[width=8.5cm,keepaspectratio]{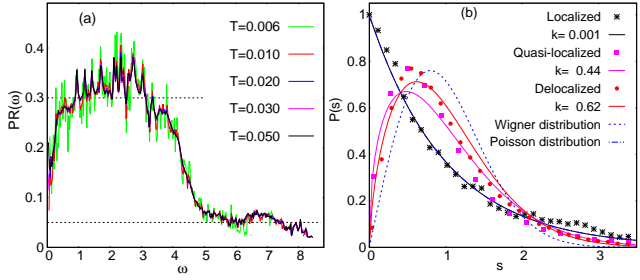}
\caption{ (Color online) (a) The average participation ratio PR$(\omega)$ as a function of $\omega$ for the quenched normal modes, same as Fig.~\ref{fig:IWM_QNM_DOS_PR}(b). The upper horizontal dotted line shows the new cut-off that demarcates the boundary between delocalized and quasi-localized modes, as discussed in the appendix~\ref{App:Lstat}. (b) Distribution $P(s)$ of the spacings, $s$, of the consecutive eigenvalues for modes having different window for participation ratio (PR) as shown in panel (a). Solid lines represent the appropriate Brody distribution associated with each type of mode with the respective Brody parameter, $k$. These illustrate the fact that an alteration of the boundary between delocalized and quasi-localized modes does not alter the qualitative conclusions of Fig.~\ref{fig:IWM_QNM_DOS_PR}(b) and Fig.~\ref{fig:IWM_QNM_LevelSpace}(a). 
}
\label{fig:IWM_Level_New}
\end{figure}

\begin{figure}[t]
\includegraphics[width=8.5cm,keepaspectratio]{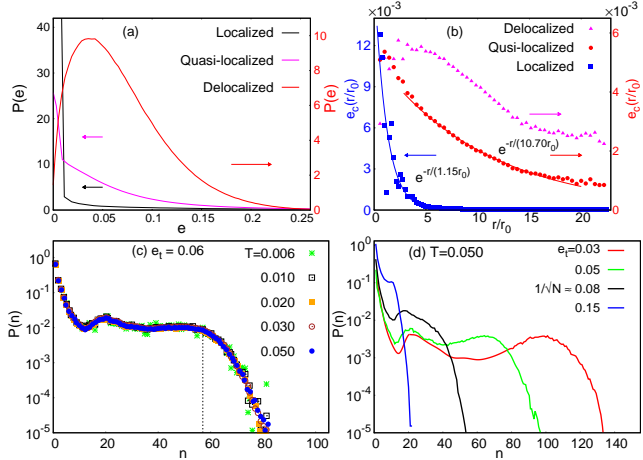}
\caption{ (Color online) (a) Distribution $P(e)$ of the magnitude $e$ of the polarization vectors for localized, quasi-localized and delocalized modes obtained using the new cut-off values of PR, as discussed in the appendix~\ref{App:Lstat}, at $T=0.050$. For localized modes, the distribution is the same as that in Fig.~\ref{fig:IWM_QNM_LevelSpace}(b).
(b) The $r$-dependence of the correlation $e_c(r)$ between the magnitude of the polarization vectors of pairs of particles separated by distance $r$, for the localized, quasi-localized and delocalized modes, for the same range of PR as in panel (a). Thick dots represent the actual data points while solid lines are exponential fits to the data points. 
(c) Probability  distribution $P(n)$ of cluster size $n$ for various $T$ in a semi-logarithmic plot with the cut-off, $e_t = 0.06$, for the magnitude of the polarization of vector, $e$, for the quasi-localized modes (with the new cut-off range of PR). (d) Probability  distribution $P(n)$ of cluster size $n$ for different values of $e_t$ at $T = 0.050$ in a semi-logarithmic plot.
}
\label{fig:IWM_New}
\end{figure}
At any given temperature $T$, $h_b(0)$ and $\dot{h}_b(0)$ can be estimated using equipartition theorem. Under harmonic approximation, we know that $\langle h_b^2 \rangle = \frac{k_B T}{\omega_b^2}$ and $\langle \dot{h}_b^2 \rangle = k_B T$. So, the typical values for the projection of displacement $(h_b)$ and velocity $(\dot{h}_b)$ on the normal mode eigenvector for a mode having frequency $\omega_b$ are $\sqrt{k_B T/\omega_b^2}$ and $\sqrt{k_B T}$, respectively. Thus, by knowing $\{\vec{r}_i^{~0}\}$, $h_b(0)$ and $\dot{h}_b(0)$, we can compute the position of the particles at any later time $t$, under the harmonic approximation, using Eq.~\ref{Eq:Harmnc_Approx}.

\begin{figure}[t]
\includegraphics[width=8.5cm,keepaspectratio]{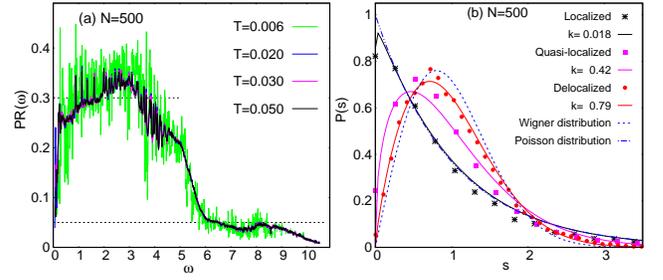}
\caption{ (Color online) (a) The average participation ratio PR$(\omega)$ as a function of $\omega$ for the quenched normal modes for $N=500$ particles. The upper horizontal dotted line shows the new cut-off that demarcates the boundary between delocalized and quasi-localized modes, as discussed in the appendix~\ref{App:Lstat}. (b) Distribution $P(s)$ of the spacings, $s$, between consecutive eigenvalues for modes $(N=500)$ having different windows for the participation ratio (PR) as shown in panel (a), at $T=0.05$. Solid lines represent the appropriate Brody distribution associated with each type of mode with the respective Brody parameter, $k$.
}
\label{fig:IWM_New_2}
\end{figure}
\section{Effects of changing the cut-off values of the participation ratio and the system size} \label{App:Lstat}

We discussed results in the main text considering quenched modes with ${\rm PR} \ge 0.35$ to be delocalized. The choice of this cut-off had an ad hoc basis. We wish to demonstrate in this appendix that tuning this threshold by a reasonable margin might change some quantitative estimates, but our key conclusions remain unaltered. For this purpose, we consider here the modes with ${\rm PR} \ge 0.30$ as delocalized (see Fig.~\ref{fig:IWM_Level_New}(a)). This amounts to $14$\% reduction of the threshold value (compared to the results in the main text) of PR to qualify modes as extended. Expectedly, this modification degrades the ``quality" of delocalized modes, and we examine below the extent of such changes, both quantitatively as well as qualitatively, on different quantities discussed in the main text.

With the new choice of the cut-off, the degree $k$ of the level spacing distribution $P(s)$ changes according to: (a) $k = 0.62$ for the delocalized modes (see Fig.~\ref{fig:IWM_Level_New}(b)), instead of $k = 0.85$ for the earlier cut-off ${\rm PR}=0.35$. (b) Similarly, for the quasi-localized modes, we find $k= 0.44$ with this new cut-off on PR, whereas, $k= 0.50$ for these modes with the cut-off used in the main text. We note that a true delocalized mode should yield $k=1$ for the Wigner distribution~\cite{RevModPhys.53.385}. Hence, with this altered cut-off, the delocalized modes depart further from a true Wigner distribution, though the nature of $P(s)$ hardly suffers any qualitative changes.

Fig.~\ref{fig:IWM_New}(a) depicts the effect of the new choice of the cut-off on the distribution $P(e)$ of the magnitude $e$ of the polarization vectors for localized, quasi-localized and delocalized modes at $T=0.050$. While the value of $e$ for which $P(e)$ becomes maximum for the delocalized modes shifts slightly towards a lower value (compared to that in Fig.~\ref{fig:IWM_QNM_LevelSpace}(b)), the nature of the distributions for different modes remains the same as that in Fig.~\ref{fig:IWM_QNM_LevelSpace}(b). Similarly, the $r$-dependence of the correlation $e_c(r)$ between the magnitude of the polarization vectors of two particles separated by a distance $r$ for different types of modes (Fig.~\ref{fig:IWM_New}(b)) and the probability  distribution $P(n)$ of the cluster size $n$ for various $T$ (Fig.~\ref{fig:IWM_New}(c)) and different $e_t$ (Fig.~\ref{fig:IWM_New}(d)) do not exhibit any significant qualitative change compared to the results described in the main text. 

While in the main text, we present results for systems with $N=150$ particles, we have verified that the key conclusions survive also for systems with $N=500$ particles. This is explicitly shown in Fig.~\ref{fig:IWM_New_2}(a-b). Fig.~\ref{fig:IWM_New_2}(a) shows the average participation ratio PR$(\omega)$ as a function of $\omega$ for the quenched normal modes for $N=500$ particles at different $T$. Distribution $P(s)$ of the spacings, $s$, of the consecutive eigenvalues for modes having the same window for participation ratio (PR) as shown in panel (a) is shown in Fig.~\ref{fig:IWM_New_2}(b) and we have identified the localized $(k=0.018)$, quasi-localized $(k=0.42)$ and delocalized $(k=0.79)$ modes even for $N=500$. However, we emphasize that increasing system size typically weakens signature of glassiness in our confinements, as discussed in an earlier publication~\cite{PRE17}. This happens because a large fraction of particles (with increasing $N$) located near the central region of the trap make up a perfect triangular lattice at low $T$, which weakens the extent of glassiness found for $N=150$. The triangular lattice of central particles remains undistorted due to a large distance from the irregular boundary, causing weakening of `disorder strength' and hence of glassiness.

\bibliography{Draft}
\bibliographystyle{apsrev}

\end{document}